\documentclass[11pt]{article}
\usepackage{algorithm}
\usepackage{graphicx,float}
\usepackage{t1enc}
\usepackage{pdfsync}
\usepackage[latin1]{inputenc}
\usepackage{mathrsfs,xspace,theorem}
\usepackage{amsmath,amssymb,bm}

\advance\textwidth by 1cm
\advance\oddsidemargin by -.5cm
\advance\evensidemargin by -.5cm
 
\advance\textheight by 2cm
\advance\topmargin by -1cm

\floatstyle{ruled}
\newfloat{Algorithm}{htpb}{alg}
\newtheorem{proposition}{Proposition}
\theorembodyfont{\rmfamily}

\floatstyle{ruled}

\def\dotminus{\mathbin{\buildrel{\hbox{.}}\over{\smash{-}\vphantom{_2}}}}

\newcommand{\shiftrs}{\mathbin{\mathalpha\gg^+}}
\newcommand{\shiftr}{\mathbin\gg}
\newcommand{\shiftl}{\mathbin\ll}
\newcommand{\band}{\mathbin\&}
\newcommand{\bor}{\mathbin|}
\newcommand{\bxor}{\oplus}
\newcommand{\bnot}[1]{\overline{#1}}
\newcommand{\hex}[1]{\mathrm{0x#1}}
\renewcommand{\div}{\mathbin\backslash}

\newcounter{prgline}
\newcommand{\pl}{\theprgline\addtocounter{prgline}{1}}

\def\..{\,\mathpunct{\ldotp\ldotp}} 

\renewcommand{\epsilon}{\varepsilon}
\renewcommand{\phi}{\varphi}

\title{Broadword Implementation of Parenthesis Queries}

\author{Sebastiano Vigna\\
\normalsize Dipartimento di Scienze dell'Informazione\\\normalsize Universit\`a
degli Studi di Milano, Italy}
\date{}
\begin{document}
\bibliographystyle{alpha}

\maketitle

\begin{abstract}
We continue the line of research started in~\cite{VigBIRSQ} proposing 
\emph{broadword} (a.k.a. SWAR---``SIMD Within A Register'') algorithms for
finding matching closed parentheses and the $k$-th far closed
parenthesis. Our algorithms work in time $O(\log w)$ on a word of $w$ bits,
and contain no branch and no test instruction. On 64-bit (and
wider) architectures, these algorithms make it possible to avoid costly
tabulations, while providing a very significant speedup with respect to
for-loop implementations.
\end{abstract}

\section{Introduction}

A \emph{succinct} data structure (e.g., a succinct tree) provides the same
operations of its standard counterpart (and sometimes more), but occupies space
that is asymptotically near to the information-theoretical lower bound. A
classical example is the $(2n+1)$-bit representation of a binary tree with $n$
internal nodes proposed by Jacobson~\cite{JacSSTG}. Recent years have
witnessed a growing interest in succinct data structures, mainly because of the
explosive growth of information in various types of text indexes (e.g., large 
XML trees).

In this paper we discuss practical implementations of two basic building
blocks: given a string of $w$ bits, where $w$ is the machine word, representing
open ($1$) and closed ($0)$ parentheses, we are interested in solving the following two
problems:
\begin{itemize}
  \item assuming the first bit is a one, finding the matching closed
  parenthesis;
  \item finding the $k$-th far closed parenthesis in the string
  (a parenthesis is \emph{far} if its matching parenthesis is not in the
  string).
\end{itemize}
Trivial solutions require scanning the string in $O(w)$ time. For the
necessities of data structures supporting operations on balanced parenthesis,
usually representing trees (see, e.g.,~\cite{JacSSTG,MuRSRBPST,JSSUSROT,GRRSORBP}), the two
operations can be implemented by tables that in principle use $o(n)$
bits for a structure with $n$ parentheses. However, the tables are actually very big, unless $n$ is very large, and they do not
usually fit the processor cache.

In this paper we push further the work started in~\cite{VigBIRSQ}, where we
argued that on modern 64-bit architecture a much more efficient approach uses
\emph{broadword programming}. The term ``broadword'' has been introduced by Don Knuth in the fascicle on
bitwise manipulation techniques of the fourth volume of \emph{The Art of
Computer Programming}~\cite{KnuACPBTT}. Broadword programming uses large (say, more than 64-bit
wide) registers as small parallel computers, processing several pieces of
information at a time. An alternative, more traditional name for similar
techniques is SWAR (``SIMD Within A Register''), a term coined 
by Fisher and Dietz~\cite{FiDCSWR}. One of the first techniques for
manipulating several bytes in parallel were actually proposed by
Lamport~\cite{LamMBPFI}. The famous HAKMEM memo~\cite{BGSH} contains several
examples of broadword programming.

We are also very careful of avoiding tests whenever possible. Branching is a
very expensive operation that disrupts speculative execution, and should be
avoided when possible. All broadword algorithms we discuss contain no test
and no branching.

While broadword programming and careful consideration of testing and cache
side-effects are by now quite common in practical implementations of succinct
data structures (see, e.g.,~\cite{DRELSTR}), to the best of our knowledge no
one has proposed broadword algorithms for the problems we study. See~\cite{GogBCFCDUCSA}
for other applications of the same ideas.

We concentrate on $64$-bit and wider architecture, but we cast all our
algorithms in a $64$-bit framework to avoid excessive notation: the modification
for wider registers are trivial. We have in mind modern processors (in
particular, the very common Opteron processor) in which
multiplications are extremely fast (actually, because the clock is slowed down
in favour of multicores), so we use them occasionally. They can be safely
replaced by $O(\log w)$ basic operation, but in practice experiments show that
on the Opteron replacing multiplications by shifts and additions, 
\emph{even in very small number}, is not competitive.

The C++/Java code implementing all data structures in this paper is available
under the terms of the GNU Lesser General Public License at
\texttt{\small http://sux.dsi.umimi.it/}.

\section{Notation}

Consider a string $\bm s$ of $n$ bits numbered from $0$. We write $s_i$
for the bit of index $i$. When can view $\bm s$ as a string of parentheses by
stipulating that 1 represent an open parenthesis, and 0 a closed parenthesis.
We define the \emph{closed excess function}
\[
E_{\bm s}(i) = |\{s_j\mid j < i \land s_j = 0\}| - |\{s_j\mid j < i \land s_j =
1\}|,
\]
which represent the excess of closed w.r.t.~open parentheses at position
$i$ (excluded). The string $\bm s$ is \emph{balanced} if the excess function is
always negative, except for $0$ and $n$, where it is zero.

We use $a\div b$ to denote integer division of $a$ by $b$, $\shiftr$ and
$\shiftl$ to denote right and left (zero-filled) shifting, $\shiftrs$ denotes
right shifting with sign extension, $\band$, $\bor$ and $\bxor$ to denote
bit-by-bit not, and, or, and xor; $\bnot x$ denotes the bit-by-bit complement of $x$. We pervasively use precedence to avoid
excessive parentheses, and we use the same precedence conventions of the
C programming language: arithmetic operators come first, ordered in the standard
way, followed by shifts, followed by logical operators; $\oplus$ sits between
$\bor$ and $\band$.

We use $L_k$ to denote the constant whose ones are in position $0$, $k$,
$2k$, \ldots\, that is, the constant with the \emph{lowest} bit of each $k$-bit subword
set (e.g, $L_8=0x01010101010101010101$). This constant is
very useful both to spread values (e.g., $\hex{12} *
L_8=\hex{1212121212121212}$) and
to sum them up, as it generates cumulative sums of $k$-bit subwords if the
values
contained in each $k$-bit subword, when added, do not exceed $k$ bits. 
(e.g., $\hex{030702}*L_8=\hex{30A0C0C0C0C0C0C0902}$---look carefully at the
 three rightmost bytes). 
We use $H_k$ to denote $L_k \shiftl k-1$, that is, 
the constant with the \emph{highest} bit of each $k$-bit subword
set (e.g, $H_8=0x8080808080808080$).

We use the notation \[
\mu_k := \bigl( 2^{2^w} - 1\bigr) \setminus \bigl( 2^{2^k} + 1\bigr), 
\]
where $\setminus$ denotes integer division.
More intuitively, $\mu_0 = \hex{5555\dots 5555}$, $\mu_1 = \hex{3333\dots
3333}$, $\mu_1 = \hex{0F0F\dots
0F0F}$, $\mu_2 = \hex{00FF\dots
00FF}$, and so on.

Our model is a RAM machine with $w$-bit words that performs logic operations,
additions and subtractions in unit time using 2-complement
arithmetic. In our algorithms we also use a constant number of multiplications,
which can be substituted with $O(\log w)$ shifts and adds without altering the
running time.

\section{Basic operations}

We recall the expression for computing in parallel the differences modulo $2^k$
of each $k$-bit subword (see~\cite{KnuACPBTT}):
\[
x -_k y := ( (x \bor H_k ) - ( y\band \bnot{H_k} ) )\bxor ( ( x \bxor \bnot y )
\band H_k ).
\]
If we know in advance that the blocks in $x$ and $y$ contain positive entries,
this simplifies to 
\[
( (x \bor H_k ) - y )\bxor H_k.
\]
Another important operation we will use is blockwise nonzero test:
\[
x\neq_k0 := \Bigl(\bigl(\, ( (x \bor H_k ) - L_k )\bor
x\bigr) \Bigr)\band H_k.
\]
Finally, truncated difference of positive entries:
\[
x \dotminus_k y = ( x -_k y ) \band ( ( x -_k y ) \shiftr k - 1 ) -_k 1 
\]
The subexpression after the $\band$ is simply a mask that cancels out every
block in which a negative result was obtained. The common subexpression $x-_k
y$ should be, of course, computed just once.

\section{Matching open parentheses}

Assume we have a string $\bm s$ such that $s_0=1$. We would like find the
associated matching closed parenthesis, if it lies in $\bm s$, or get some
special value otherwise. 
The general strategy to obtain this result in $O(\log w)$ time and $O(1)$
additional space is to consider the excess function, as 
clearly we are interested in computing
\[
\min_{0\leq j<w} E_{\bm s}(j) = 0.
\]
We operate in the following manner: we will \emph{sample} $E_{\bm s}$ each
$2^{\lceil\log\log w\rceil}$ positions. Then, we will scan
linearly in parallel each of the resulting $w/2^{\lceil\log\log w\rceil}$
blocks from the end, recording whether in some block the function crosses zero,
and where this happens. Finally, we find the first block that hit a zero and return
the corresponding position.

Let us first consider a 64-bit sampling phase on input $x$; blocks are just bytes this case. We start with a small variant of the
standard broadword algorithm for sideways additions:
\begin{tabbing}
\setcounter{prgline}{0}
\hspace{0.3cm} \= \hspace{0.3cm} \= \hspace{0.3cm} \= \hspace{0.3cm} \=
\hspace{0.3cm} \=\kill\\
\pl\>$b = x - ( x \band \hex{AAAAAAAAAAAAAAAA} ) \shiftr 1$\\
\pl\>$b = ( b \band \hex{3333333333333333} ) + ( ( b \shiftr 2 ) \band
\hex{3333333333333333} )$\\ 
\pl\>$b = ( b + ( b \shiftr 4 ) ) \band \hex{0F0F0F0F0F0F0F0F0}$\\
\pl\>$b = ( b * L_8 ) \shiftl 1$
\end{tabbing}
At this point, each byte of $b$ contains \emph{twice} the number of
open parentheses appearing up to that block, included. Note that the excess
function satisfies 
\[
E_{\bm s}(j) = |\{s_j\mid j < i \land s_j = 0\}| - |\{s_j\mid j < i \land s_j =
1\}| = j - 2|\{s_j\mid j < i \land s_j =
1\}|,
\]
so getting a sample of $E_{\bm s}$ each $8$ bits just requires parallel
subtraction with a suitable constant:
\[b = ( H_8 \bor \hex{4038302820181008}) -_8 b.\]
Note the presence of $H_8$, which avoid propagation of the sign bit, and in
practice let us represent each sample in two's complement in the seven lower
bits of each byte. We now set up an \emph{update mask} $u$ that contains, for each
byte of $b$, zero, if the byte is nonzero in the lower seven bits, 
but $\hex{7F}$ otherwise: \[u = ((((b \bor H_8 ) -L_8) \shiftr 7 \band L_8 ) |
H_8 ) - L_8\]
 
Using $u$ we set up our last variable $z$, that throughout the computation 
will contain, for each byte of $b$, either zero, if the byte was never equal to
zero (in the lower seven bits), or a counter expressing the position of the
parenthesis that caused the excess function to go to zero. If we find a zero
byte initially, the position is clearly 7:
\[
z = ( H_8 \shiftr 1 \bor L_8 * 7 ) \band u.
\]
We now update $b$, modifying the values of the excess function \emph{two bits
at a time}: this is correct, as a balanced string has necessarily even length,
so the excess function cannot go to zero at an odd position. In the first round
we thus compute
\[
b = b - ( L_8 *2 - ( ( x \shiftr 6 \band L_8 \shiftl 1 ) + ( x \shiftr 5 \band
L_8 \shiftl 1)))
\]
We now recompute $u$ as above, but update $z$ as follows:
\[
z = z \band \bnot u \bor ( H_8 \shiftr 1 \bor L_8 * 5 ) \band u.
\]
Due to the update rule, even nonzero bytes of $z$ will be updated. This is
correct, as we want to find the zero of the excess function that is closer to
the first bit. We continue in this way until we have completed scanning each
byte: the next update of $b$ is thus
\[
b = b - ( L_8 *2 - ( ( x \shiftr 4\band L_8 \shiftl 1 ) + ( x \shiftr 3 \band
L_8 \shiftl 1))),
\]
and so on. Finally, we gather our result by locating the relevant block using an
LSB operator (e.g., Brodal's~\cite{KnuACPBTT}), which we assume to return $-1$ in
case no bit is set:
\begin{tabbing}
\setcounter{prgline}{0}
\hspace{0.3cm} \= \hspace{0.3cm} \= \hspace{0.3cm} \= \hspace{0.3cm} \=
\hspace{0.3cm} \=\kill\\
\pl\>$p = \operatorname{LSB}(z \shiftr 6 \band L_8)$\\
\pl\>$( ( p + ( z \shiftr p \band \hex{3F} ) )\bor ( p \shiftr 8 ) ) ) \band
\hex{7F}$
\end{tabbing}
The last line contains the expression returned (we will return $127$ in case no
matching parenthesis exists).

The algorithm is best followed on an example: consider the first two bytes
of a 64-bit string:
\[
\framebox{1 1 0 0 1 0 1 1}\quad \framebox{0 0 0 0 1 0 1 0} \ldots
\]
Note that the left most bit is bit zero. We are representing the string of
parentheses
\[
\framebox{( ( ) ) ( ) ( (} \quad\framebox{) ) ) ) ( ) ( )} \ldots
\]
The excess function behaves as follows:
\begin{center}
\includegraphics[scale=.7]{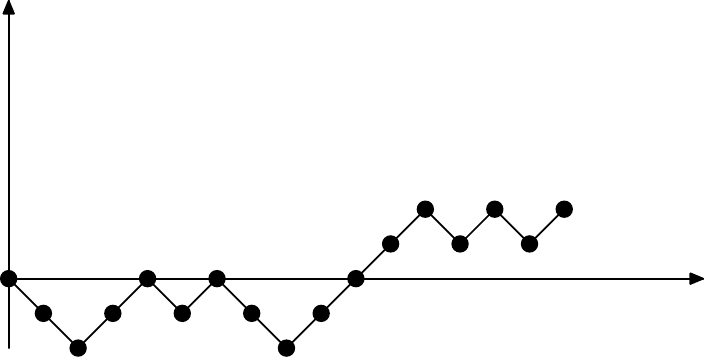}
\end{center}
In the first computation step, we sample the excess function at each byte, so
the first bytes of $b$ (in two's complement) are -2 and +2. No result is
thus stored in $z$. However, in the first update we modify the samples of the excess function by subtracting
the contribution of the underlined parentheses:
\[
\framebox{1 1 0 0 1 0 \underline1 \underline1}\quad \framebox{0 0 0 0 1 0
\underline1 \underline0}
\ldots
\]
Now the first byte of $b$ changes to 0, so we store in $z$ our result as
follows:
\[
\framebox{0 1 0 0 0 1 0 1}\quad \framebox{0 0 0 0 0 0 0 0} \ldots
\]
The sixth bit records that there is a value, and for the time being the
candidate result is $5$. Note that the current result
is spurious, because there is another zero to be found.

We now update again $b$, subtracting another pair of parentheses:
\[
\framebox{1 1 0 0 \underline1 \underline0 \underline1 \underline1}\quad
\framebox{0 0 0 0 \underline1 \underline0 \underline1 \underline0}
\ldots
\]
The first byte of $b$ is now $0$, the second byte $+2$. Thus, $z$ is
updated as follows:
\[
\framebox{0 1 0 0 0 0 1 1}\quad \framebox{0 0 0 0 0 0 0 0} \ldots
\]
In the last update, the \emph{second} byte of $b$ becomes $0$. The final
value of $z$ is thus
\[
\framebox{0 1 0 0 0 0 1 1}\quad \framebox{0 1 0 0 0 0 0 1} \ldots
\]
Now the LSB operator detects that the first zero is in the first byte, and the
correct value (3) is extracted from $z$ and returned.

The construction of the first sample requires $O(\log\log w)$ instructions and
a single multiplication (which could be substituted with $O(\log w)$
operations). The parallel linear scan clearly requires $O(\log w)$ operations.

\section{Finding far closed parentheses by index}

In this section we discuss the problem of finding the $p$-th
far closed parenthesis. The simple combinatorial idea at the heart of the
algorithm is the following, easily proved statement:

\begin{proposition}
\label{prop:far}
Let $\bm t$, $\bm u$ be bit strings, $\#_{\mathrm{open}}$/$\#_{\mathrm{closed}}$ the operators returning the
number of far open/closed parenthesis in a string, and $\dotminus$
truncated subtraction. Then:
\begin{align*}
\#_{\mathrm{open}} \bm t\bm u &= (\#_{\mathrm{open}} \bm t \dotminus \#_{\mathrm{closed}} \bm u)+ \#_{\mathrm{open}} \bm u\\
\#_{\mathrm{closed}} \bm t\bm u &= (\#_{\mathrm{closed}} \bm u \dotminus
\#_{\mathrm{open}} \bm t)+ \#_{\mathrm{closed}} \bm u
\end{align*}
\end{proposition}

Since it is easy to compute the number of far open/closed parenthesis in 2-bit
blocks using masking, and it is also easy to do parallel truncated subtraction,
using additional $O(\log w)$ words we can compute the number of far open/closed
parentheses in blocks of length $2^i$, $2\leq i<\log w$. At that point, we can
use the above property backwards: if we are searching from the $k$-th far
closed parenthesis in $\bm t\bm u$, this must be either in $\bm t$, if
$k <\#_{\mathrm{closed}}\bm t$, or in $\bm u$, but in position
$k-\#_{\mathrm{closed}}\bm t+ \#_{\mathrm{open}}\bm t$. We will assume for the
time being that a $p$-th far closed parenthesis \emph{does} exist in the string. Results will be unpredictable otherwise.

In general, each $2^k$-bit block of the variables $o_k$ and $c_k$ will keep
track of the number of far open/closed parentheses 
in the corresponding $2^k$-bit block of the input $x$. We bootstrap
our computation by filling $o_1$ and $c_1$:
\begin{tabbing}
\setcounter{prgline}{0}
\hspace{0.3cm} \= \hspace{0.3cm} \= \hspace{0.3cm} \= \hspace{0.3cm} \=
\hspace{0.3cm} \=\kill\\
\pl\>$b_0 = x \band \hex{5555555555555555}$\\
\pl\>$b_1 = ( x \band \hex{AAAAAAAAAAAAAAAA} ) \shiftr 1$\\
\pl\>$l = ( b0 \bxor b1 ) \band b1$\\
\pl\>$o_1 = ( b0 \band b1 ) \shiftl 1 \bor l$\\
\pl\>$c_1 = ( ( b0 \bor b1 ) \bxor \hex{5555555555555555} ) \shiftl 1 \bor l$
\end{tabbing}
These operations implements the mappings
\begin{align*}
 00 &\to 00 &00 &\to 10\\
 01 & \to 01& 01& \to 01 \\
 10 & \to 00&10 &\to 00\\
 11 & \to 10& 11 &\to 00 
 \end{align*}
They send each 2-bit substring to the number of far open, or closed,
respectively, parentheses.

The $k$-th phase, $1<k<\log w$, records in temporary variables $e_o$ and $e_c$
the number of far open and far closed parentheses in each half of $2^{k+1}$-bit
blocks. These numbers are then combined using Proposition~\ref{prop:far}:
\begin{tabbing}
\setcounter{prgline}{0}
\hspace{0.3cm} \= \hspace{0.3cm} \= \hspace{0.3cm} \= \hspace{0.3cm} \=
\hspace{0.3cm} \=\kill\\
\pl\>$e_o = o_k \band \mu_k$\\
\pl\>$e_c = ( c_k \band \mu_k \shiftl 2^k ) \shiftr 2^k$\\
\pl\>$o_{k+1} = ( ( o_k \band \mu_k \shiftl 2^k ) \shiftr 2^k ) + ( e_o
\dotminus_8 e_c )$\\ \pl\>$c_{k+1} = ( c_k \band \mu_k ) + ( e_c \dotminus_8 e_o
)$\
\end{tabbing}
Finally, we work backwards, isolating the part of the string containing the
required parenthesis. At the $k$-th step, $k = \log w - 1, \log w -2, \dots, 1$
we operate as follows:
\begin{tabbing}
\setcounter{prgline}{0}
\hspace{0.3cm} \= \hspace{0.3cm} \= \hspace{0.3cm} \= \hspace{0.3cm} \=
\hspace{0.3cm} \=\kill\\
\pl\>$b = (( p - ( c_k \shiftr s \band 2^{2^k} -1 ) ) \shiftrs w - 1 )-
1$\\ \pl\>$m = b \band  2^{2^k} -1 $\\
\pl\>$p \mathrel{\mathalpha{-}\mathalpha{=}} c_k \band m$\\
\pl\>$p \mathrel{\mathalpha{+}\mathalpha{=}} o_k \band m$\\
\pl\>$s \mathrel{\mathalpha{+}\mathalpha{=}} 2^k \band b$\\
\end{tabbing}
The variable $s$ keeps track of the left (i.e., lowest) extreme of the
interval of width $2^{k+1}$ in which we are performing our binary search.
Initially, $s$ is zero and $k=\log w - 1$, which means that we are searching
for the $p$-th far closed parenthesis in the whole string $\bm s$.

In each phase, we first of all set $b$ so that it is $0$ if the $p$-th
far closed parenthesis appears in the block of length $2^k$ starting at position $s$, 0 otherwise. 
Note that we can do this because the far closed parentheses in the \emph{first}
half are true, global far closed parentheses. We
then set up our mask $m$, which will be used to update $p$: if $m$ is zero,
there is no update to do---we just have to restrict our search interval.
Otherwise, we have to decrease $p$ by $c_k\band m$ (as we are skipping $c_k
\band m$ far closed parentheses) and increase it by $o_k\band m$ (as there are
$o_k\band m$ far open parentheses before the block we're moving in, so we must
offset $p$). Finally, $s$ must be updated and moved forward by $2^k$ in case
$b\neq 0$.

In the last phase, we are left with a two-bits string and a value $p$. It is
easy to check that the following hand-crafted expression gives the correct
result:
\[
s + p + ( ( x \shiftr s \band ( ( p \shiftl 1 ) \bor 1 ) ) \shiftl 1 ).
\]
Finally, it is easy to see that be performing an additional phase in the
first part of the algorithm we can obtain the overall number of far
closed parentheses in the whole string, making it easy to return a special
value in case the requested parenthesis does not exist.

\section{Experiments}
\label{sec:exp}

We performed a number of experiments on a Linux-based system sporting a 64-bit
Opteron processor running at $2814.501$\,MHz with $1$\,MiB of first-level cache. The tests show that
on 64-bit architectures broadword programming provides significant performance
improvements. We compiled using \texttt{gcc} 4.1.2 and options \texttt{-O9}.

Our previous experience with similar code shows that testing in isolation very
tight code can produce paradoxical results. It is much more informative to
embed the code in a typical simple application: in our case, we implemented
Jacobson's classical $O(n)$ balanced parentheses representation~\cite{JacSSTG}
and performed tests measuring the time required to find a matching closed
parenthesis using our broadword algorithms and a tuned for-loop implementation.

The experimental setting for benchmarking operations that require
nanoseconds must be set up carefully. We generate at random bit arrays
containing correctly parenthesised strings, and store a
million test positions. During the tests, the positions are read with a
linear scan, producing minimal interference; generating random positions
during the tests causes instead a significant perturbation of the results,
mainly due to the slowness of the modulo operator. 
The tests are repeated ten times and averaged. We measure user time using the
system function \texttt{getrusage()}.

Generating random balanced strings of parenthesis requires some attention. We
use Arnold and Sleep's classical algorithm~\cite{ArSURGBPS}, but with a
\emph{twist}. The algorithm chooses at each step whether to add a closed
parenthesis with probability
\[
P_{r,k}=\frac12\frac{r(k+r+2)}{k(r+1)},
\]
where $r$ is the number of open parentheses still to be closed, and $k$ the
remaining number of symbols to be generated. Note that when $k=r$ we
have $P_{r,k}=1$, so we just generate closed parentheses.

To estimate better the behaviour of our algorithms, we introduce a
\emph{twist}, that is, a number $0\leq t\leq 1$ that shifts the probability so
that open parentheses are more likely to be generated. In other words,
\[
P_{r,k,t} =\begin{cases}
                  1 & \text{if $P_{r,k}=1$;}\\
                  tP_{r,k} & \text{otherwise.}
                  \end{cases}
\]
The result is that when $t<1$ we will tend to generate strings with deeper
nesting. We are interested in experimenting with the behaviour at different
deepness levels because trivial (for-loop) solutions behave very well on random 
strings because most open parentheses are near, and moreover their matching
parenthesis is a few bits away. But if you consider a typical application, for
instance, binary search trees, then a search going down into a large tree has to
find a far matching parenthesis for most of the search. More precisely, it is
not difficult to see that for a complete binary tree the \emph{average} (over
all paths going from the root to a leaf) distance between the open and closed
parenthesis of a query is $\Theta(n/\log n)$ (assuming the binary tree is
mapped to a forest using the inverse of the
first-child/next-sibling isomorpshism, and that the forest is represented using balanced
parentheses in the standard way). To simulate this fact, we use a
skewed distribution: we plan to enlarge, however, our test set with more realistic large search trees or XML trees.

\begin{figure}
\centering
\includegraphics{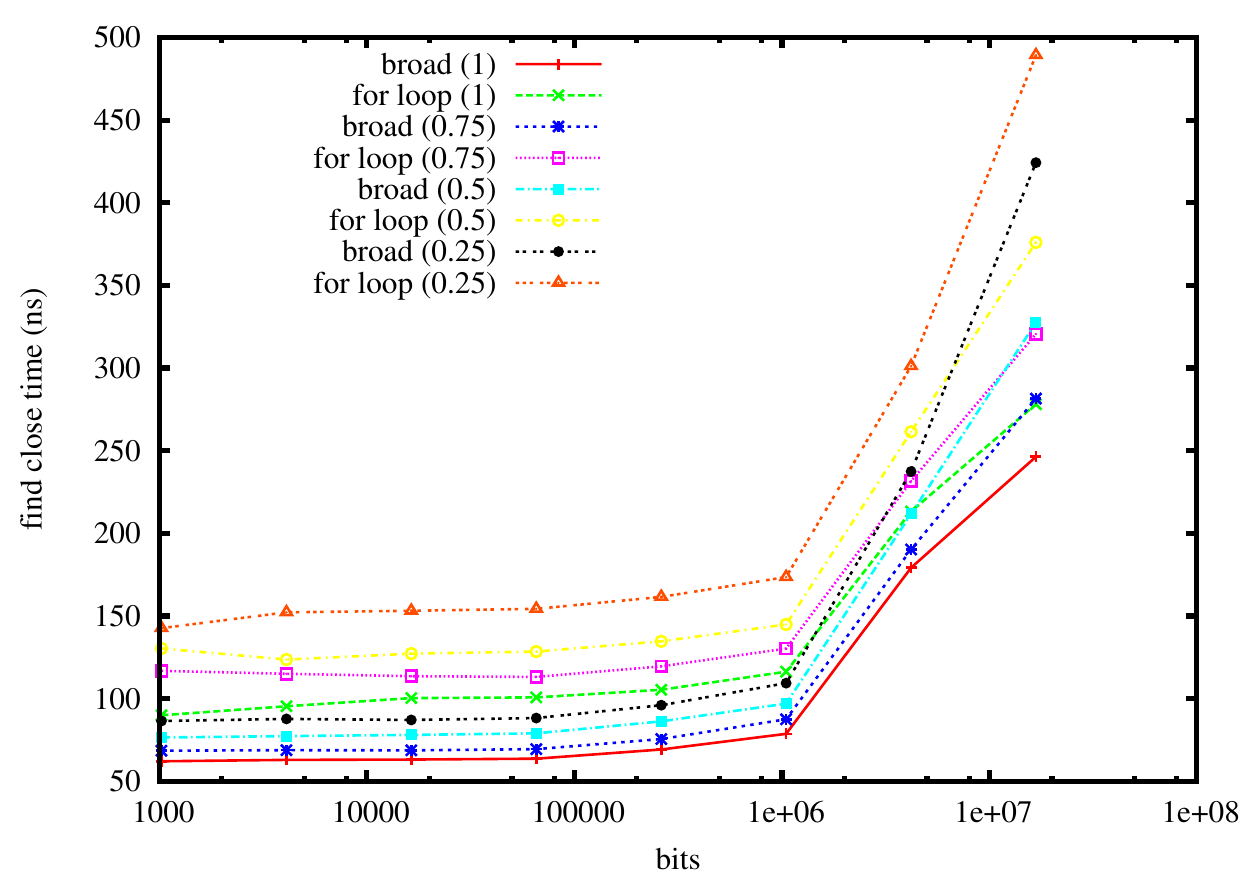}
\caption{\label{fig:exp}A graph displaying the data shown in
Table~\ref{tab:exp}. Up to around one million bit the timings remain constant even in practice; after that,
memory access becomes significant and size has a significant effect on speed
(as in the case of rank/select queries---see~\cite{GGMPIRSQ}).}
\end{figure}

We compare our structures against tuned for-loop implementations: results 
are shown in Table~\ref{tab:exp} and Figure~\ref{fig:exp}, which clearly show
the advantage of the broadword implementation, in particular for longer matchings (e.g., for low
twist). We expect, of course, that figures will improve as $w$ gets larger.

\begin{table}
\centering
\begin{tabular}{r||r|r|r|r|r}
& 1 & .75 & .50 & .25 \\
\hline

1\,Ki&62.10/89.90&68.50/116.80&76.50/130.40&86.50/142.60\\
4\,Ki&62.90/95.40&68.80/115.00&77.30/123.60&87.70/152.20\\
16\,Ki&63.10/100.30&68.70/113.50&78.00/127.20&87.10/153.20\\
64\,Ki&63.70/100.70&69.40/113.10&79.00/128.50&88.20/154.30\\
256\,Ki&69.20/105.40&75.50/119.50&86.20/134.70&96.00/161.60\\
1\,Mi&78.70/116.30&87.50/130.30&97.00/144.90&109.30/173.50\\
4\,Mi&179.20/213.20&190.20/231.60&211.80/261.50&237.40/301.20\\
16\,Mi&246.30/278.20&281.50/320.60&327.50/376.10&424.30/489.30
\end{tabular}
\caption{\label{tab:exp}Timings in nanoseconds for a parenthesis matching
operation in Jacobson's data structure. The first value is obtained used the
broadword algorithms presented in this paper, whereas the second value is obtained using a for-loop
implementation. Column labels show the amount of twisting,
whereas row labels show the number of parentheses in the string.}
\end{table}

\section{Conclusions}

Extending some previous work of ours~\cite{VigBIRSQ}, we
have introduced some two new broadword algorithms that implement
two basic operations typical of succinct static
data structures for balanced parentheses.
We have also presented experiments that 
compares our results with a for-loop baseline. We discussed our algorithms in
the case of closed parentheses, but they can be immediately modified to find
matching open or far open parentheses.

We leave for future work experimentation with tabulated implementations.
The latter tend to be, of course, very fast when tested, but they engage the
processor cache significantly, and their global impact cannot be measured
easily. For-loop implementations have a cache footprint similar to that of our
broadword versions, so they are first natural candidate for comparisons.

A Java version of this code is currently distributed by the Sux4J
project\footnote{\texttt{\small http://sux4j.dsi.unimi.it/}.} as part of a
highly compressed implementation of a monotone minimal perfect hash function
(see~\cite{BBPTPMMPH}).

\small

\bibliography{biblio}

\end{document}